\documentclass[twocolumn,aps,prd,superscriptaddress,showpacs]{revtex4}

\usepackage{color} 

\usepackage{amsmath}
\usepackage{amssymb}
\usepackage{amsfonts}
\usepackage{amsthm}

\usepackage{latexsym}
\usepackage{mathrsfs}
\usepackage[cmtip]{xypic}
\usepackage[all]{xy}
\usepackage{appendix}
\usepackage[pdftex]{graphicx}
\usepackage{braket}

\newcommand{\wt}{\widetilde}

\newcommand{\la}{\langle}
\newcommand{\ra}{\rangle}

\begin{document}

\title{The Configuration Space of Conformal Connection-Dynamics}

\author{James A. Reid} 
\email[]{j.a.reid@abdn.ac.uk}
\affiliation{SUPA Department of Physics, University of Aberdeen, King's College, Aberdeen, AB24 3UE, UK} 
\author{Charles H.-T. Wang}
\email[]{c.wang@abdn.ac.uk}
\affiliation{SUPA Department of Physics, University of Aberdeen, King's College, Aberdeen, AB24 3UE, UK}
\affiliation{STFC Rutherford Appleton Laboratory, Chilton, Didcot, Oxon, OX11 0QX, UK}

\date{August 17, 2012}

\begin{abstract}
The configuration space of the reduced Hamiltonian formulation of quantum gravity has been shown, for non-Ricci flat metrics, to be a higher-dimensional analogue of the Teichm\"{u}ller space of conformal structures on a Riemann surface. In this article we show that the configuration space of conformal connection-dynamics is naturally a higher-dimensional Teichm\"{u}ller space, subject to the same condition.  An immediate consequence of this result is that the Barbero-Immirzi parameter of loop quantum gravity naturally assumes a dilatonic character in all conformal canonical gravity theories.  
\end{abstract}


\maketitle

\section{Introduction}
Over $50$ years, the structure of the configuration space of canonical gravity has successively evolved in sophisticatication.  The mathematical structures in terms of which it is defined have undergone a similar evolution - from the original space of Riemannian 3-metrics in the ADM formulation of general relativity as a constrained Hamiltonian system \cite{adm} to a higher-dimensional analogue of Teichm\"uller space in Fischer \& Moncrief's reduced Hamiltonian formulation of quantum gravity \cite{fischer}.\\ \\
Motivated by consonances with gauge theory, Ashtekar in the late $1980$'s reformulated canonical gravity in terms a new set of variables \cite{ashtekar1}: a soldering form $\sigma^{a \phantom{A} B}_{\phantom{a} A}$ and a spin connection, $A_{aA}^{\phantom{aA}B}$. The new theory was called \emph{connection-dynamics}, wherein the metric was no longer a fundamental variable, but rather a composite object.  When Ashtekar formulated his connection-dynamics, the contempraneous understanding of the configuration space of canonical gravity, so called \emph{superspace}, was the space of all positive-definite Riemannian 3-metrics modulo diffeomorphisms: $\{ \left[ g \right] \}$, as developed by Fischer \cite{fischersuperspace} and Ebin \cite{ebin}.  The configuration space of canonical gravity in spinorial variables was defined in an analogous manner \cite{ashtekarbook}, as the space of all (suitably-regular) soldering forms modulo diffeomorphisms: $\{ \left[ \sigma \right] \}$.  By writing the metric as a product of two soldering forms, it was clear by extension that the two configuration spaces were $2$-$1$ homomorphic (as moduli spaces.)\\ \\
In light of the pioneering work of Deser \cite{deser} and later York \cite{york2} on the covariant decomposition of symmetric tensors; Fischer \& Moncrief - in an attempt to reduce Einstein's equations to an unconstrained Hamiltonian system - refined the structure of superspace \cite{fischer2}.  For non-Ricci flat metrics, they identified it with a higher-dimensional analogue of \emph{Teichm\"uller space}, which in complex analysis is the moduli space of conformal structures on a Riemann surface.  In this article, we show that when one formulates canonical gravity as conformal connection-dynamics in the sense of \cite{charlesconn} or \cite{charles1}, subject to the same non-Ricci flatness condition, one naturally finds the configuration space to be a Teichm\"{u}ller space.  This has important physical ramifications as we will discuss shortly.\\ \\
The structure of the paper is as follows.  We begin in section \ref{geometrodynamicssection} by introducing some concepts from modern conformal geometry, and show that the existence of a conformal structure on a spatial hypersurface naturally partitions tensor fields into conformal classes.  We then discuss Fischer \& Moncrief's construction of Teichm\"{u}ller space in section \ref{superspacesection}, and show in section \ref{conformalspinsection} that there exist conformal spin structures on spatial hypersurfaces, which are necessary to partition the spinorial basic variables into conformal classes - just as a conformal metric structure is necessary to partition world tensors into conformal classes.  It is then shown that the configuration space of conformal connection-dynamics is a Teichm\"{u}ller space in the sense of Fischer \& Moncrief \cite{fischer2}.  Schematically, we show that the diagram
\begin{equation}
\xymatrix{
 \{ \left[ \sigma \right] \} \ar[r] \ar[d]_{\psi_{\text{old}}} & \ar[d]^{\psi_{\text{new}}} \{ \left[ \langle \sigma  \rangle \right] \} \\
  \{ \left[ g \right] \} \ar[r] & \{ \left[ \langle g \rangle \right] \}
}
\end{equation}
commutes, where $\{ \left[ \langle g \rangle \right] \}$ and $\{ \left[ \langle \sigma  \rangle \right] \}$ are the Teichm\"{u}ller spaces.  In section \ref{barberosection} we show that this result has immediate ramifications for the Barbero-Immirzi parameter of loop quantum gravity \cite{immirzi}.  Various comments abound in canonical gravity literature concerning the role of the Barbero-Immirzi as a scalar field (see \cite{taveras}-\cite{torres} for example), and was first suggested by one of us in \cite{charlesconn}.  Whilst the dilatonic character of the Barbero-Immirzi parameter had been found by imposing an additional constraint to generate conformal diffeomorphisms \cite{charles1}, we use our formalism to show that such a character is generic to all conformal canonical gravity theories by virtue of the underlying conformal geometry.

\section{Conformal Geometry}\label{geometrodynamicssection}
In this section we introduce some salient concepts from conformal geometry which will be required throughout.  In particular, we work towards showing that conformal transformations partition tensors into conformal classes.  To do so, we introduce some mathematical objects which are essentially absent in conformal physics (see \cite{gover} for one of the few examples), but are ubiquitious in mathematics (see \cite{bailey}-\cite{armstrong}).  These are called conformal weight bundles, and since they may be defined for conformal spin structures as well as conformal metric structures, we use them to show that the configuration space of conformal connection-dynamics is a Teichm\"{u}ller space.   For reviews of the modern approach to conformal geometry, we refer the reader to \cite{eastwood}, \cite{baumjuhl} \& \cite{leitner}.\\ \\
Let $M$ denote an $n$-dimensional manifold and $g$ a Riemannian metric on $M$.  We say that a \emph{conformal structure} \cite{alekseevskii} on $M$ is a class $C$ of pairwise-homothetic metrics 
\begin{equation}
C = \{ \lambda g~|~\lambda \in \mathbb{R}^{+} \}.
\end{equation}
The presence of a conformal structure induces a bundle of conformal frames over the manifold \cite{baumjuhl}.  To formalise this notion, and set notation, let $\pi: P_{G}(M) \to M$ denote a principal fibre bundle with structure group $G$ on a manifold $M$ and let us denote by $\Gamma(P_{G}(M))$ its space of sections.  When $G$ is the \emph{linear conformal group}
\begin{equation}\label{linearconfgroup}
CO(n) := SO(n) \times \mathbb{R}^{+}
\end{equation}
on $M$, we say that the principal $CO(n)$-bundle
\begin{equation}
\pi: P_{CO(n)}(M) \to M
\end{equation}
is the \emph{conformal frame bundle}, where a \emph{conformal frame $\theta_{c}(p)$} at $p \in M$ is a pseudo-orthonormal basis in $T_{p}M$, with respect to some fixed metric $g$.   Indeed, the standard fibre of the conformal frame bundle is the set of all conformal frames at some $p \in M$.  Importantly, the 1-dimensional center of the conformal group allows one to define a class of weighted density line bundles, $\mathcal{E}[w]$.  In the notation of \emph{associated} fibre bundles (see \cite{isham} for an introduction), these are defined by
\begin{align}\label{weightbundles}
\mathcal{E}[w] :&= P_{CO(n)}(M) \times_{\rho} \mathbb{R}\nonumber \\
\rho(c)(x) &= -w \frac{\det (\theta_{c})}{2n} x,
\end{align}
for a defining representation $\rho$, a conformal frame $\theta_{c}$ and $x \in \mathbb{R}$. These are the so-called \emph{conformal weight bundles}, which conformally weight sections of bundles associated to the frame bundle and tensor powers thereof.  Their particulars are best introduced by way of example.  Let us regard the cotangent bundle on some $n$-manifold $M$ as an associated bundle:
\begin{equation}
T^{*}M := P_{SO(n)} \times_{\rho} \left( \mathbb{R}^{n} \right)^{*}.
\end{equation}
Under a conformal transformation $\phi$, a metric $g$ is conformally weighted according to
\begin{equation}
\phi: g_{ab} \mapsto \Omega^{\frac{4}{n-2}} g_{ab}
\end{equation}
where the power $4/(n-2)$ is conventional in canonical gravity \cite{choquetbruhat}. Examining this at the level of spaces of sections of bundles makes the role of the conformal weight bundles in the weighting more transparent
\begin{equation}
\phi: \Gamma(\odot^{2} T^{*}M) \to \Gamma \left( \mathcal{E} \left[ \frac{4}{n-2} \right] \otimes \odot^{2} T^{*}M \right).
\end{equation}
For clarity, let us adopt Penrose's abstract index notation, where the bundles to which a certain tensor is a section are themselves adorned with indices.  For example, a 2-form $t_{ab} \in \Gamma(\mathcal{E}[w] \otimes \Lambda^{2} T^{*}M))$ of conformal weight $w$ would simply be written $t_{[ab]} \in \mathcal{E}_{[ab]}[w]$.  This notation at hand, it may be seen that a conformal transformation induces the partitioning of general tensors into conformal classes.  Consider the conformal transformation $\phi$ of some tensor $t_{a \ldots b}^{c \ldots d}$ of conformal weight $w$.  $\phi$ acts as
\begin{equation}\label{eq:phiaction}
\xymatrix{\phi : \Gamma \left( \mathcal{E}_{a \ldots b}^{c \ldots d} \right) \ar[r] & \ar@{=}[d] \Gamma \left( \mathcal{E}_{a \ldots b}^{c \ldots d}[w] \right). \\
& \Gamma \left( \mathcal{E} [w] \otimes \mathcal{E}_{a \ldots b}^{c \ldots d} \right)
}
\end{equation}
We see from \eqref{eq:phiaction} clear that $\phi$ maps $t$ to the set $\{ \Omega^{w} t ~|~ \Omega^{w} \in \Gamma(\mathcal{E}[w]) \}$, which we term its \emph{conformal class}, $\la t \ra$.  Indeed, a conformal transformation may be viewed as an automorphism of the conformal class, just as multiplying two functions together may be viewed as an automorphism of a group of functions.

\section{Superspace}\label{superspacesection}
This notation in place, we now examine Fischer \& Moncrief's identification of the configuration space of the reduced Hamiltonian formulation of quantum gravity with a higher-dimensional Teichm\"{u}ller space.  Our presentation is a condensed review of that given in \cite{fischer}.  Let us denote by $\mathcal{F}^{+}$ the infinite-dimensional Abelian group of positive functions on some $n$-manifold $M$, acting on the space of Riemannian 3-metrics, $\mathcal{M}$, by pointwise multiplication:
\begin{equation}\label{fischerstart}
\mathcal{F}^{+} \times \mathcal{M} \to \mathcal{M}.
\end{equation}
The orbit of $\mathcal{F}^{+}$ through $g$ is identified with $\langle g \rangle$, and we term the resulting orbit space,
\begin{equation}
\mathcal{M} / \mathcal{F}^{+} = \{ \langle \emph{g} \rangle ~|~\emph{g} \in \mathcal{M} \},
\end{equation}
the space of \emph{conformal classes on} $M$. Fischer \& Moncrief define an $\mathcal{F}^{+}$-principal fibre bundle over this space by projection:
\begin{align}\label{projection}
\pi_{\mathcal{M}} :~ \mathcal{M} &\to \mathcal{M} / \mathcal{F}^{+} \nonumber \\
 g &\mapsto \langle g \rangle.
\end{align}
We take the diffeomorphism group $\mathcal{D}$ of $M$ to act on $\mathcal{M} / \mathcal{F}^{+}$ according to the commutative diagram:
\begin{equation}
\xymatrix{\mathcal{M} \times \mathcal{D} \ar[d] \ar[r]^{f} & \mathcal{M} \ar[d]\\
\mathcal{M} / \mathcal{F}^{+} \times \mathcal{D} \ar[r]_{f^{*}} & \mathcal{M}/\mathcal{F}^{+}}
\end{equation}
where the vertical arrows are the projections of the $\mathcal{F}^{+}$-principal fibre bundle $\mathcal{M} \to \mathcal{M} / \mathcal{F}^{+}$, and the pullback of a diffeomorphism $d$ along $\pi_{\mathcal{M}}$ is given by
\begin{equation}
f^{*} : (\langle g \rangle, d) \mapsto \langle d^* g \rangle
\end{equation}
for some conformal class $\la g \ra$.  Now, for a particular $\langle g \rangle \in \mathcal{M} / \mathcal{F}^{+}$,
\begin{equation}
\left[ \langle g \rangle \right] = \{d^* \langle g \rangle ~|~ d \in \mathcal{D} \}
\end{equation}
is the orbit of the diffeomorphism group through $\la g \ra$, so that $\left[ \langle g \rangle \right]$ is itself a class of diffeomorphically-equivalent conformal classes.   In canonical gravity, the resulting orbit space
\begin{equation}\label{conformalsuperspace}
\frac{\mathcal{M} / \mathcal{F}^{+}}{\mathcal{D}} = \{ \left[ \langle g \rangle \right] ~|~ \langle g \rangle \in \mathcal{M} / \mathcal{F}^{+} \}  
\end{equation}
is traditionally called \emph{conformal superspace} \cite{barbouromurchadha}, which Fischer \& Moncrief identified with a higher-dimensional analogue of Teichm\"{u}ller space for certain metrics.  To preface the identification, let $\mathcal{D}_{0}$ denote the connected component of the identity of $\mathcal{D}$.  The quotient space
\begin{equation}\label{fischerfinish}
\mathcal{T} \equiv \frac{\mathcal{M} / \mathcal{F}^{+}}{\mathcal{D}_{0}}
\end{equation}
is precisely the (higher-dimensional) \emph{Teichm\"uller space} of conformal structures on $M$.  It was found that such an association is only valid for those spatial hypersurfaces $M \in \mathcal{M}$ which are of Yamabe type $-1$, which is to say, those that admit no Ricci-flat metrics.  In general, Fischer \& Moncrief remark, neither does $\mathcal{D}_{0}$ act freely on $\mathcal{M} / \mathcal{F}^{+}$ nor is Teichm\"uller Space a manifold \cite{fischer}.  There is evidence however, that these issues may - at least in a minisuperspace approach - be resolvable \cite{reid_twistorstructures}.

\section{Conformal Spin Structure}\label{conformalspinsection}
As will be defined shortly, the basic dynamical variables of Ashtekar's reformulation of canonical gravity are $SU(2)$ soldering forms.  These are innately spinorial in nature, and so their partitioning into conformal classes is conditional on the existence of a conformal spin structure on a spatial hypersurface.  One cannot \emph{apriori} partition soldering forms into conformal classes using the conformal metric structure, since 3-metrics are no longer fundamental entities in the theory, but rather are composed of two soldering forms.  It is for this reason that we must show that there exist spatial hypersurfaces $M \in \mathcal{M}$ which are endowed with conformal spin structures.\\ \\
To introduce these, let $P_{SO(n)}(M) \to M$ be the orthonormal frame bundle over an $n$-manifold, $M$. $P_{SO(n)}(M)$ is, by definition, an $SO(n)$-principal fibre bundle and we have from \cite{ammann} that a \emph{spin structure} on $M$ is a pair $(P_{\text{Spin}(n)}(M), \vartheta)$, where $P_{\text{Spin}(n)}(M) \to M$ is a $\text{Spin}(n)$-principal fibre bundle and $\vartheta : P_{\text{Spin}(n)}(M) \to P_{SO(n)}(M)$ is a $\mathbb{Z}_{2}$ quotient such that the diagram
\begin{equation}\label{eq:conformalspinstructure}
\xymatrix{P_{\text{Spin}(n)}(M) \ar[dd]^{\vartheta \times \Theta} \ar[r] \times \text{Spin}(n) & P_{\text{Spin}(n)}(M) \ar[dd]_{\vartheta} \ar[dr] &  \\
& & M  \\
P_{SO(n)}(M) \times SO(n) \ar[r] &  P_{SO(n)}(M) \ar[ur] &}
\end{equation}
commutes, and where
\begin{equation}\label{eq:spinnson}
\Theta : \text{Spin}(n) \to SO(n)
\end{equation}
is the (non-trivial) double cover of $SO(n)$.  We propose the following:\\ \\
\textbf{Proposition (Existence of Spin Structure)}.
There exist orientable 3-manifolds $M \in \mathcal{M}$ representing spatial hypersurfaces in a Lorentzian spacetime for which the diagram
\begin{equation}\label{eq:conformalspinstructure_particular}
\xymatrix{P_{SU(2)}(M) \ar[dd]^{\vartheta \times \Theta} \ar[r] \times SU(2) & P_{SU(2)}(M) \ar[dd]_{\vartheta} \ar[dr] &  \\
& & M  \\
P_{SO(3)}(M) \times SO(3) \ar[r] &  P_{SO(3)}(M) \ar[ur] &}
\end{equation}
commutes and thereby constitutes a spin structure.\\ \\
\textbf{Proof (Existence of Spin Structure)}.
Considering the orthonormal frame bundle $P_{SO(3)}(M)$, we have that since $M$ is orientable, $P_{SO(3)}(M)$ is parallelizable. The parallelizability of the orthonormal frame bundle allows the global construction of the $\text{Spin}(3)$-principal bundle $P_{\text{Spin}(3)}(M)$, and the principal fibre bundle morphism $\vartheta: P_{\text{Spin}(3)}(M) \to P_{SO(3)}(M)$ is a $\mathbb{Z}_{2}$ quotient for those $M$ whose second Stiefel-Whitney class vanishes.\\ \\
Now, the theory of simple Lie Algebras tells us that
\begin{equation}\label{spinnson}
\text{Spin}(3) \cong SU(2),
\end{equation}
deriving from the isomorphism $A_{1} \cong B_{1}$ of the root systems of the simple Lie algebras $A_{1}$ and $B_{1}$.  Inserting this into \eqref{eq:spinnson} we recover the well known fact that $SU(2)$ is the non-trivial double cover of $SO(3)$, and so $\Theta: \text{Spin}(3) \to SO(3)$. In light of this, we write $P_{\text{Spin}(3)} \to M$ as $P_{SU(2)}(M) \to M$ .  Substituting these facts into the commutative diagram \eqref{eq:conformalspinstructure}, we recover the proposition \eqref{eq:conformalspinstructure_particular}.\\ \\
Having shown that $\vartheta$ and $\Theta$ are well defined, as is their product
\begin{equation}
\vartheta \times \Theta : P_{SU(2)}(M) \times SU(2) \longrightarrow P_{SO(3)}(M) \times SO(3)
\end{equation}
by projection onto respective factors.\\ \\
Lastly, the horizontal arrows $P_{SU(2)}(M) \times SU(2) \to P_{SU(2)}(M)$ and $P_{SO(3)}(M) \times SO(3) \to P_{SO(3)}(M)$ in \eqref{eq:conformalspinstructure} are right $G$-actions, and so \eqref{eq:conformalspinstructure} commutes as proposed. \hfill $\square$\\ \\
Parenthetically, it was worth remarking that those $M \in \mathcal{M}$ which do not admit spin structures are precisely those with non-vanishing second Stiefel-Whitney classes, $H^{2}(M, \mathbb{Z}_{2}) > 0$.  In this case, $P_{SO(3)}(M)$ is not a $\mathbb{Z}_{2}$ quotient of $P_{\text{Spin}(3)}(M)$, and so $\vartheta$ is not realised.\\ \\
Now, a foundational result in conformal spin geometry states that a conformal spin structure exists if and only if a spin structure exists \cite{ammann}.  From a fibre bundle point of view this is natural, since a conformal structure on a generic fibre bundle $E \to M$ may be taken to be a sub-bundle
\begin{equation}
\begin{array}{ll}
\pi : & C \to M \\
&\cap\\
&E
\end{array}
\end{equation}
of the bundle $E \to M$ whose fibres $\pi^{-1} (p) = C_{p}$ are conformal structures on the fibres $E_{p}$ of $E \to M$.  In essence, one may pass to $C$ by restriction.\\ \\
Let us now give the definition of a conformal spin structure.  Recall from (\ref{linearconfgroup}) that $CO(n) := SO(n) \times \mathbb{R}^{+}$ is the linear conformal group, the structure group of the conformal frame bundle $\pi: P_{CO(n)}(M) \to M$.  It is a well known fact in spin geometry \cite{ammann} that the double cover (\ref{spinnson}) extends to a double cover
\begin{equation}\label{cspinncson}
\xymatrix{C\text{Spin}(n) \ar@{=}[d] \ar[r]^{\Theta_{C} \times \text{id}} & CO(n). \\
\text{Spin}(n) \times \mathbb{R}^{+} &}
\end{equation}
So, analogous to the definition of a spin structure, we say that a \emph{conformal spin structure} is given by a $C\text{Spin}(n)$-principal fibre bundle, $P_{C\text{Spin}(n)}(M) \to M$ together with a map $\vartheta_{C} : P_{C\text{Spin}(n)}(M) \to P_{CO(n)}(M)$ such that the commutative diagram
\begin{equation}\label{css}
\xymatrix{
P_{C\text{Spin}(n)}(M) \ar[dd]^{\vartheta_{C} \times \Theta_{C}}  \ar[r] \times C\text{Spin}(n) & P_{C\text{Spin}(n)}(M) \ar[dd]_{\vartheta_{C}} \ar[dr] &  \\
& & M  \\
P_{CO(n)}(M)\times CSO(n) \ar[r] &  P_{CO(n)}(M)\ar[ur] &
}
\end{equation}
commutes.  Notice from (\ref{cspinncson}) that the conformal spin group $C\text{Spin}(n)$ has a nontrivial 1-dimensional center which gives rise to a class of conformal lines bundles by using a spin representation in equation (\ref{weightbundles}).  To distinguish them from the conformal line bundles arising from the conformal metric structure, we write them as $\mathcal{E}[w]_{S}$.  Recall from section \ref{geometrodynamicssection} that the endowment of conformal weights under conformal transformation extended to sections of (tensor powers of) bundles associated to the frame bundle, as is the case here.\\ \\
Let us now construct an associated bundle to the principal $SU(2)$-bundle of spin frames, $P_{SU(2)}(M)$, and examine the effects of a conformal transformation on its sections. The \emph{spinor bundle} on $M$ is defined as:
\begin{equation}
S := P_{SU(2)}(M) \times_{\rho} \text{End}(\Sigma)
\end{equation}
where $\text{End}(\Sigma)$ is the space of endomorphisms of a two complex-dimensional vector space $\Sigma$.  The abstract notation hides the fact that we are dealing with SU(2) spinors, but one may see from \cite{belgun} that in three dimensions: the two fundamental representations $\text{End}(\Sigma)$ and $\text{End}(\Sigma')$ of the Clifford algebra bundle $\text{Cl}(M)$ on $M$ are isomorphic to the fundamental representation of $\text{Spin}(3) \subset \text{Cl}(M)$, and so both are in fact isomorphic to SU(2).\\ \\
Now, let us examine the effect of a conformal transformation on a composite spinor.  By the usual convention, we write spinorial indices with upper-case Latin letters, and so $\psi_{A \ldots B} \in \mathcal{E}_{A \ldots B}$, for example.  Under a conformal transformation
\begin{equation}\label{eq:phiactionspin}
\xymatrix{\phi : \Gamma \left( \mathcal{E}_{A \ldots B} \right) \ar[r] & \ar@{=}[d] \Gamma \left( \mathcal{E}_{A \ldots B}[w]_{S} \right), \\
& \Gamma \left( \mathcal{E}[w]_{S} \otimes \mathcal{E}_{A \ldots B} \right)
}
\end{equation}
and so the spinors $\psi_{A \ldots B}$ are partitioned into conformal classes according to $\psi \mapsto \{ \Omega^{w} \psi ~|~ \Omega^{w} \in \Gamma(\mathcal{E}[w]_{S}) \}$, in the same manner as world-indexed tensors.  This in mind, let us consider the basic variables of connection-dynamics, the $SU(2)$ soldering forms.\\ \\
An $SU(2)$ \emph{soldering form}, $\sigma$, is a 1-form on the bundle of spin frames with values in $S \otimes S^{*}$.  Soldering forms are examples of \emph{generalized tensors} \cite{ashtekartensor}, meaning that they possess both world- and spinorial indices: $\sigma^{a \phantom{A} B}_{\phantom{a} A}$.  As basic variables, they occur in canonical gravity \emph{densitized} by the square root of the determinant of the 3-metric
\begin{equation}
\wt{\sigma}^{a \phantom{A} B}_{\phantom{a} A} = \sqrt{g}~ \sigma^{a \phantom{A} B}_{\phantom{a} A}
\end{equation}
Given their world and spinorial indices, these are naturally partitioned by \eqref{eq:phiaction} and \eqref{eq:phiactionspin} according to
\begin{equation}\label{spart}
\la \wt{\sigma} \ra := \{ \Omega^{w} \wt{\sigma} ~|~ \Omega^{w} \in \mathcal{F}^{+} \},
\end{equation}
where $w$ is a net conformal weight and we regard the conformal factors $\Omega^{w}$ as positive functions for greater utility.  To see that the configuration space of canonical gravity in the spinorial variables is a Teichm\"{u}ller space, consider the space of densitized soldering forms $\mathcal{S} := \{ \wt{\sigma} \}$, with the indices suppressed for brevity.   Fischer \& Moncrief's procedure of constructing the higher-dimensional Teichm\"{u}ller space follows exactly precisely as in (\ref{fischerstart})$-$(\ref{fischerfinish}), and so one may write the configuration space of conformal connection-dynamics as
\begin{equation}
\mathcal{T}_{S} = \frac{\mathcal{S}/ \mathcal{F}^{+}}{\mathcal{D}_{0}},
\end{equation}
for those soldering forms whose products are non-Ricci flat metrics.  Writing the product explicitly:
\begin{equation}\label{metricconstruction}
g^{ab} = - \wt{\sigma}^{a \phantom{A} B}_{\phantom{a} A} \wt{\sigma}^{b \phantom{B} A}_{\phantom{b} B},
\end{equation}
it is clear that $\psi_{\text{new}} : \mathcal{T}_{S} \to \mathcal{T}$ is a double cover of moduli spaces.  In light of this result, work in preparation suggests that the cotangent bundle of Teichm\"{u}ller space may - in a certain minisuperspace approach - be free of both non-manifold points \& conical singularities and have the structure of a local twistor bundle \cite{reid_twistorstructures}, illuminating a new aspect of conformal geometry in canonical gravity.

\section{Barbero-Immirzi Parameter}\label{barberosection}
We shall now examine the ramifications that the partitioning of soldering forms into conformal classes (\ref{spart}) has for the Barbero-Immirzi ambiguity of loop quantum gravity.  This ambiguity arises from the fact that the spectra of the quantum geometrical observables - such as area or volume - are not uniquely determined by a fixed pair of classical variables \cite{mercuri}. The resulting one-parameter families of unitarily-inequivalent quantum theories are parameterised by the Barbero-Immirzi parameter, $\beta$.\\ \\
Sections $E^{i}_{\phantom{i}a}$ of the orthonormal frame bundle $P_{SO(3)}(M) \to M$ are termed \emph{triads} in this context.  From \cite{peres}, one may densitize the triad according to
\begin{equation}
\wt{E}^{a}_{\phantom{a}i} = \epsilon^{abc} \epsilon_{ijk} E^{j}_{\phantom{j}b} E^{k}_{\phantom{k}c},
\end{equation}
where the $\epsilon$'s are alternating tensors. The \emph{Barbero-Immirzi parameter} is defined \cite{immirzi} as the pre-factor $\beta$ on the extrinsic curvature term in the \emph{Barbero connection}
\begin{equation}
A^{(\beta)i}_{\phantom{i}a} := \Gamma^{i}_{\phantom{i}a} + \beta K^{i}_{\phantom{i}a},
\end{equation}
where $\Gamma^{i}_{\phantom{i}a}$ is the Levi-\v{C}ivit\`{a} connection and $K^{i}_{\phantom{i}a} =  ( \text{det}~ \wt{E}^{i}_{\phantom{i}a} )^{-\frac{1}{2}} K_{ab} E^{k}_{\phantom{k}j} \delta^{ij}$ the extrinsic curvature.  We wish to find an expression for the extrinsic curvature in terms of soldering forms, and so we note from \cite{ashtekarbook} that the latter are expressible in terms of the triads as
\begin{equation}
\sigma^{a \phantom{A} B}_{\phantom{a} A} = -\frac{i}{\sqrt{2}} E^{a}_{\phantom{a}i}  \tau^{i\phantom{A}B}_{\phantom{i}A\phantom{B}}, 
\end{equation}
where $\tau^{i\phantom{A}B}_{\phantom{i}A\phantom{B}}$ are the Pauli matrices.  Thus, the triads are expressible as
\begin{equation}
E^{a}_{\phantom{a}i} = \sqrt{2} i \sigma^{a \phantom{A} B}_{\phantom{a} A} \tau_{iB}^{\phantom{i}\phantom{B}A},
\end{equation}
and so we are able to compute the extrinsic curvature term in the Barbero Connection in terms of soldering forms:
\small{
\begin{align}\label{longcalcone}
K^{i}_{\phantom{i}a} &=  \frac{1}{\sqrt{\text{det}~ \tilde{E}^{i}_{\phantom{i}a}}} K_{ab} E^{k}_{\phantom{k}j} \delta^{ij} \nonumber \\
&=  \frac{1}{\sqrt{\text{det}~ \tilde{E}^{i}_{\phantom{i}a}}} K_{ab} \left( \frac{1}{2} \epsilon^{bcd} \epsilon_{jkl} E^{k}_{\phantom{k}c} E^{l}_{\phantom{l}d} \right) \delta^{ij} \nonumber \\
&= - \frac{1}{\sqrt{\text{det}~ \tilde{E}^{i}_{\phantom{i}a}}} K_{ab} \epsilon^{bcd} \epsilon_{jkl} \sigma_{cA}^{\phantom{c}\phantom{A}B} \sigma_{dC}^{\phantom{d}\phantom{C}D} \tau^{k\phantom{B}A}_{\phantom{k}B} \tau^{l\phantom{D}C}_{\phantom{l}D} \delta^{ij}.
\end{align}
\normalsize
Recall now that $\sigma_{cA}^{\phantom{c}\phantom{A}B}$ is only unique up to a conformal factor, since $\sigma_{cA}^{\phantom{c}\phantom{A}B}$ and $\sigma_{dC}^{\phantom{d}\phantom{C}D}$ are fixed representatives of their conformal class in conformal connection-dynamics.  To highlight the generality of the ambiguity, upon conformal transformation (\ref{longcalcone}) becomes
\begin{equation}
K^{i}_{\phantom{i}a} = - \frac{\Omega^{w}}{\sqrt{\text{det}~ \tilde{E}^{i}_{\phantom{i}a}}} K_{ab} \epsilon^{bcd} \epsilon_{jkl} \sigma_{cA}^{\phantom{c}\phantom{A}B} \sigma_{dC}^{\phantom{d}\phantom{C}D} \tau^{k\phantom{B}A}_{\phantom{k}B} \tau^{l\phantom{D}C}_{\phantom{l}D} \delta^{ij}.
\end{equation}
where $w$ is the combined conformal weight of the product of the (undensitized) soldering forms $\sigma$.  Now, since the group of positive functions $\mathcal{F}^{+}$ is closed under scaling by an arbitrary positive number, there exists some conformal factor $\omega \in \mathcal{F}^{+}$ such that
\begin{equation}
\varphi = \beta \Omega^{w}
\end{equation}
where $\beta$ is a positive real number.  This restricts the cases where conformal symmetry may `absorb' the Immirzi parameter into the conformal factor, the former being, by definition, a complex parameter.  Continuing with the calculation,
\small
\begin{align}
A^{(\beta)i}_{\phantom{i}a} &= \Gamma^{i}_{\phantom{i}a} + \beta K^{i}_{\phantom{i}a} \nonumber \\
&= \Gamma^{i}_{\phantom{i}a} - \frac{\beta \Omega^{w}}{\sqrt{\text{det}~ \tilde{E}^{i}_{\phantom{i}a}}} K_{ab} \epsilon^{bcd} \epsilon_{jkl} \sigma_{cA}^{\phantom{c}\phantom{A}B} \sigma_{dC}^{\phantom{d}\phantom{C}D} \tau^{k\phantom{B}A}_{\phantom{k}B} \tau^{l\phantom{D}C}_{\phantom{l}D} \delta^{ij} \nonumber \\
&=  \Gamma^{i}_{\phantom{i}a} - \frac{\varphi}{\sqrt{\text{det}~ \tilde{E}^{i}_{\phantom{i}a}}} K_{ab} \epsilon^{bcd} \epsilon_{jkl} \sigma_{cA}^{\phantom{c}\phantom{A}B} \sigma_{dC}^{\phantom{d}\phantom{C}D} \tau^{k\phantom{B}A}_{\phantom{k}B} \tau^{l\phantom{D}C}_{\phantom{l}D} \delta^{ij}\nonumber  \\
&= \Gamma^{i}_{\phantom{i}a} + \varphi K^{i}_{\phantom{i}a}.
\end{align}
\normalsize
We see immediately that provided the Barbero-Immirzi is real and positive it is transformed into a dilaton.  Whilst this behaviour had been obtained by one of us by imposing additional constraints to generate conformal diffeomorphisms \cite{charles1}, it is clear that it arises naturally from the existence of the conformal spin- and conformal metric structures which induce the partitioning of the soldering forms into conformal classes.  Therefore, this behaviour is in fact a general feature of conformal canonical gravity theories such as conformal connection-dynamics \cite{charlesconn}, \cite{charles1} and shape dynamics \cite{ggk}, \cite{gk}.

\begin{acknowledgments}

We are grateful to J. Barbour, S. Gryb, T. Koslowski and L. Smolin for formative remarks, to S. Deser for recent comments and to R. Bingham and the STFC Centre for Fundamental Physics for hospitality.  J.R. warmly acknowledges a Perimeter Scholars Internatonal scholarship from the Perimeter Institute for Theoretical Physics/University of Waterloo where this work was first envisioned and an EPSRC PhD Studentship under the grant EP/I036877/1 enabling its completion.

\end{acknowledgments}


\begin{thebibliography}{99}

\bibitem{adm}
R. Arnowitt, S. Deser and C. Misner in \emph{Gravitation: An introduction to current research}, L. Witten, ed. (Wiley, New York, 1962); arXiv:0405109 [gr-qc] (2004)

\bibitem{fischer}
A. E. Fischer and V. Moncrief in \emph{Physics on Manifolds, Proceedings of the International Colloquium in honour of Yvonne Choquet-Bruhat} (Kluwer Academic Publishers, Boston, 1992)

\bibitem{ashtekar1}
A. Ashtekar, Phys. Rev. Lett. \textbf{57}, 2244

\bibitem{fischersuperspace}
A. Fischer, in \emph{Relativity: Proceedings of the Relativity Conference in the Midwest} (Plenum Press, New York, 1970)

\bibitem{ebin}
D. Ebin, Bull. Amer. Math. Soc. \textbf{74}, 5 (1968)

\bibitem{ashtekarbook}  
A. Ashtekar and R. Tate, \emph{Lectures on Non-Perturbative Canonical Gravity} (World Scientific Publishing Ltd., 1991) 

\bibitem{deser}
S. Deser, Ann. Inst. Henri Poincar\'{e} \textbf{7}, 2 (1967)

\bibitem{york2}
J. W. York, J. Math. Phys. \textbf{14}, 456 (1973)

\bibitem{fischer2}
A. E. Fischer and V. Moncrief, Gen. Rel. Grav. \textbf{28} (1996)

\bibitem{charlesconn}
C. H.-T. Wang, Phys. Rev. D \textbf{72}, 087501 (2005)

\bibitem{charles1}
C. H.-T. Wang, Phil. Trans. R. Soc. A \textbf{66}, 1867 (2008)

\bibitem{immirzi}
G. Immirzi, Class. Q. Grav. \textbf{14}, 10 177 (1997)

\bibitem{taveras}
V. Taveras and N. Yunes,  Phys. Rev. D \textbf{78}, 064070 (2008)

\bibitem{mercuri}
S. Mercuri and V. Taveras, Phys. Rev. D \textbf{80}, 104007 (2009)

\bibitem{torres}
A. Torres and K. Krasnov, Phys. Rev. D \textbf{79}, 104014 (2009)

\bibitem{gover}
A. R. Gover, A. Shaukat and A. Waldron, Phys. Lett. B \textbf{675}, 93 (2009)

\bibitem{bailey}
T. Bailey, M. Eastwood, M. and A. R. Gover, Rocky Mountain J. Math \textbf{24}, 4 1191 (1994)

\bibitem{eastwood}
M. Eastwood in \emph{Proceedings of the 15th Winter School ``Geometry and Physics''} (Sede della Societ\`{a}, 1996)

\bibitem{armstrong}
S. Armstrong, J. Geom. Phys. \textbf{57}, 10 2024 (2007)

\bibitem{baumjuhl} 
H. Baum, and A. Juhl, A. \emph{Conformal Differential Geometry: $Q$-Curvature and Conformal Holonomy} (Birkh\"{a}user Mathematics, Basel, 2010) 

\bibitem{leitner}
F. Leitner, in \emph{Complex and Differential Geometry, Springer Proceedings in Mathematics 8} (Springer-Verlag, Berlin, 2010)

\bibitem{alekseevskii}
D. Alekseevskii in \emph{Encyclopedia of Mathematics} (Springer-Verlag, Berlin, 2011)

\bibitem{isham}
Isham, C. \emph{Modern Differential Geometry for Physicists}, World Scientific Lecture Notes in Physics

\bibitem{choquetbruhat}
Choquet-Bruhat, Y. \emph{General Relativity and the Einstein Equations} (Oxford University Press, Oxford, 2009)

\bibitem{barbouromurchadha} 
J. Barbour and N. O'Murchadha, arXiv:1009.3559 [gr-qc] (2010)

\bibitem{reid_twistorstructures}
J. A. Reid, and C. H.-T. Wang, In preparation (2012)

\bibitem{ammann}
Ammann, B. http://www.berndammann.de/publications (2003)

\bibitem{belgun}
F. Belgun, J. Geom. Phys. \textbf{37}, 229 (2001)

\bibitem{ashtekartensor}
A. Ashtekar, G. Horowitz and A. Magnon-Ashtekar, Gen. Rel. Grav. \textbf{14}, 5 (1982)

\bibitem{peres}
A. Perez,  arXiv:0409061 [gr-qc] (2005)

\bibitem{ggk}
H. Gomes, S. Gryb and T. Koslowski, Class. Q. Grav. \textbf{28}, 045005 (2011)

\bibitem{gk}
H. Gomes and T. Koslowski, Class. Q. Grav. \textbf{29}, 075009 (2012)

\end{thebibliography}
\end{document}